\documentclass[aps,twocolumn,amssymb,amsfonts,amsmath,showpacs,final,prl]{revtex4-1}

\usepackage{float}
\usepackage[dvips,final]{graphicx}
\usepackage{color}
\usepackage{amsmath}

\begin{document}

\title{Competing spin transfer and dissipation at Co/Cu(001) interfaces \\ on femtosecond timescales}

\author{J. Chen}
\affiliation{Faculty of Physics and Center for Nanointegration (CENIDE), University of Duisburg-Essen, Lotharstr.~1, 47057 Duisburg, Germany}

\author{U. Bovensiepen}
\affiliation{Faculty of Physics and Center for Nanointegration (CENIDE), University of Duisburg-Essen, Lotharstr.~1, 47057 Duisburg, Germany}

\author{A. Eschenlohr}
\email[]{andrea.eschenlohr@uni-due.de}
\affiliation{Faculty of Physics and Center for Nanointegration (CENIDE), University of Duisburg-Essen, Lotharstr.~1, 47057 Duisburg, Germany}

\author{T. M\"uller}
\affiliation{Theory Department, Max Planck Institute for Microstructure Physics, Weinberg~2, 06120 Halle, Germany}

\author{P. Elliott}
\affiliation{Theory Department, Max Planck Institute for Microstructure Physics, Weinberg~2, 06120 Halle, Germany}

\author{E. K. U. Gross}
\affiliation{Theory Department, Max Planck Institute for Microstructure Physics, Weinberg~2, 06120 Halle, Germany}

\author{J. K. Dewhurst}
\affiliation{Theory Department, Max Planck Institute for Microstructure Physics, Weinberg~2, 06120 Halle, Germany}

\author{S. Sharma}
\email[]{sangeeta.sharma@mbi-berlin.de}
\affiliation{Theory Department, Max Planck Institute for Microstructure Physics, Weinberg~2, 06120 Halle, Germany}
\affiliation{Max Born Institute for Nonlinear Optics, Max-Born-Strasse 2A, 12489 Berlin, Germany}

\date{\today}

\begin{abstract}

By combining interface-sensitive non-linear magneto-optical experiments with femtosecond time resolution and \textit{ab-initio} time-dependent density functional theory, we show that optically excited spin dynamics at Co/Cu(001) interfaces proceeds via spin-dependent charge transfer and backtransfer between Co and Cu. This ultrafast spin transfer competes with dissipation of spin angular momentum mediated by spin-orbit coupling already on sub 100~fs timescales. We thereby identify the fundamental microscopic processes during laser-induced spin transfer at a model interface for technologically relevant ferromagnetic heterostructures.

\end{abstract}

\maketitle

The interaction of femtosecond (fs) laser pulses with magnetically ordered materials leads to magnetization dynamics on femto- to picosecond (ps) timescales in a highly non-equilibrium regime. Effects like ultrafast demagnetization \cite{beaurepaire1996, kirilyuk2010}, all-optical switching \cite{kirilyuk2013, stupkiewicz2017}, and coherent control of magnons \cite{kampfrath2011, bossini2016} are highly relevant for potential future ultrafast spintronics applications. In particular, laser-driven spin-dependent charge currents on nm length scales induce fs spin transport in ferromagnetic (FM)/paramagnetic (PM) metallic heterostructures \cite{malinowski2008, battiato2010, melnikov2011, rudolf2012, turgut2013, eschenlohr2013, kampfrath2013, schellekens2014, razdolski2017, hofherr2017}, which are the building blocks for spintronic devices, and chemically inhomogeneous ferrimagnetic alloys \cite{graves2013} exhibiting all-optical switching.

However, the role of microscopic processes at the interfaces of heterostructures is far from understood. Photo-induced charge transfer which includes excitation and relaxation of charge carriers, potentially involving nuclear motion, is fundamental in such heterosystems and highly dynamic \cite{rothberg2000, nitzan, staehler2008}. Spin injection, i.e. spin-dependent charge transfer \cite{cinchetti2009, chang2015}, further becomes important at FM/PM interfaces as well as symmetry breaking leading to a modification of spin-orbit coupling (SOC). Control over these spin currents opens up a wide, previously inaccessible field of magnetization control by light pulses, e.g. ultrafast tuning of the magnetic order \cite{oistr}. However, due to the complexity of addressing spin-dependent charge transfer processes at FM/PM interfaces directly on their intrinsic time and length scales, and identifying competing local loss mechanisms of spin polarization \cite{zhang2000, toews2015, wieczorek2015, jal2017, reid2018, dornes2018} and their microscopic origin, a full control of ultrafast non-local spin dynamics remains elusive.

\begin{figure}
\includegraphics[width=\columnwidth]{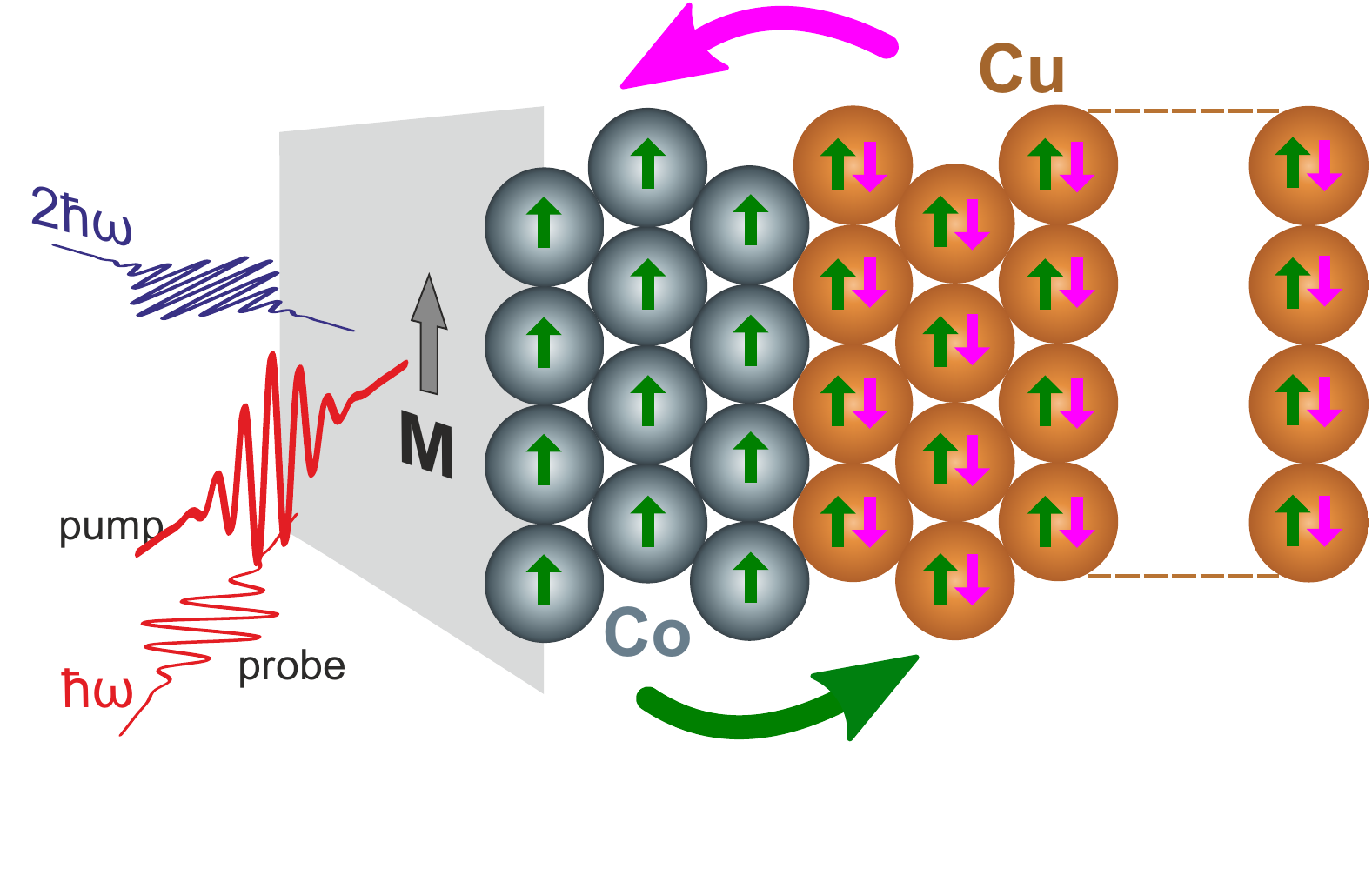}
\caption{Illustration of the epitaxial model interface Co/Cu(001), the interface sensitive pump-probe experiment, and spin transfer dynamics. Vertical arrows indicate spins and ferromagnetic order in Co. Cu carries negligible spin polarization before optical excitation. Horizontal arrows represent spin transfer across the interface. The Co magnetization $M$ is oriented perpendicular to the optical plane. Pump and probe pulses are $s$- and $p$-polarized, respectively.}
    \label{fig:1}
\end{figure}

In this Letter, we overcome this limitation by combining interface-sensitive femtosecond time-resolved non-linear magneto-optics \cite{guedde1999, regensburger2000, rasing1999, schmidt2005, melnikov2003} with \textit{ab-initio} time-dependent density functional theory calculations on the identical epitaxial interface system, namely Co/Cu(001). TDDFT provides a parameter-free description of non-equilibrium dynamics, and the underlying mechanisms are not assumed but rather emerge from the theory. In particular, we go beyond a model description based on bulk transport properties such as superdiffusive transport \cite{battiato2010}. We directly compare our \textit{ab-initio} description with experiment and are able to distinguish all relevant microscopic processes during spin injection at the Co/Cu(001) interface. We showcase the competition of photo-excited spin transfer from Co to Cu and back transfer from Cu to Co as well as demagnetization by SOC-mediated spin flips, all within the time interval shorter than 100~fs. Our results present a crucial step towards solving the critical open question of the role of the interface in ultrafast spin transport and competing elementary processes in FM heterosystems.

We become exclusively sensitive to the Co/Cu(001) interface by studying ultrathin films of 3 and 5 monolayer (ML) thickness and employing second harmonic generation (SHG) \cite{guedde1999, chen2017} as a probe, which in centrosymmetric media is generated at interfaces only, where the inversion symmetry is broken. Co/Cu(001) films with epitaxial, atomically sharp interfaces \cite{weber1996, jaehnke1999} are prepared, characterized, and measured \textit{in situ} at room temperature in ultrahigh vacuum at a pressure smaller than $10^{-10}$~mbar. In a pump-probe experiment using 35~fs (FWHM) laser pulses with 800~nm wavelength, we analyze laser-induced magnetization dynamics with SHG at 400~nm wavelength after excitation at an incident pump fluence $F=4\pm2$~mJ/cm$^2$. Our films are magnetized parallel to the sample surface, perpendicular to the optical plane, and we detect SHG of the probe pulse in transversal geometry, see figure~\ref{fig:1}. From the second harmonic (SH) intensities $I^{\uparrow,\downarrow}$ for opposite orientations of the magnetization \textbf{M} we derive the SH fields, $|E^{2\omega}_{\mathrm{even}}| \approx \sqrt{\frac{I^{\uparrow}+I^{\downarrow}}{2}}$ and $|E^{2\omega}_{\mathrm{odd}}| \approx \frac{I^{\uparrow}-I^{\downarrow}}{4|E^{2\omega}_{\mathrm{even}}|}$, which behave even and odd with reversal of \textbf{M}, respectively, and are considered magnetization-\textit{in}dependent and -dependent for $E^{2\omega}_{\mathrm{even}} >> E^{2\omega}_{\mathrm{odd}}$ \cite{guedde1999, conrad2001, note_even}. The time-dependent changes of $E^{2\omega}_{\mathrm{even,odd}}$ are normalized to their respective values before laser excitation and are represented by $\Delta^{2\omega}_{\mathrm{even,odd}}=\frac{E^{2\omega}_{\mathrm{even,odd}}(t)}{E^{2\omega}_{\mathrm{even,odd}}(t<0)}-1$, which measure charge and spin dynamics separately. For further details see \cite{chen2017}.

\begin{figure}
\includegraphics[width=\columnwidth]{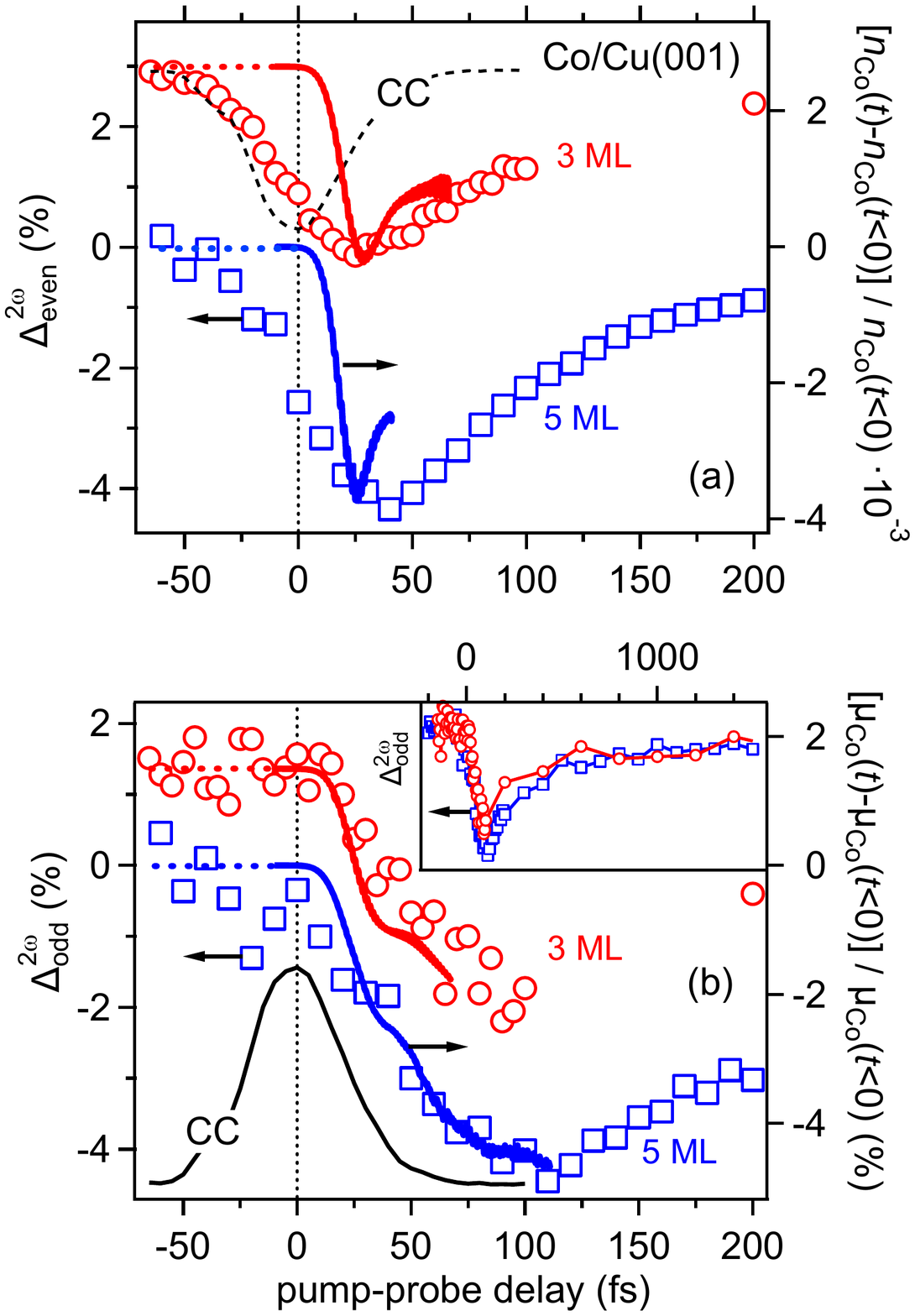}
\caption{Pump-induced relative changes in second harmonic fields $\Delta^{2\omega}_{\mathrm{even}}$ (a) and $\Delta^{2\omega}_{\mathrm{odd}}$ (b) for 3 (circles) and 5~ML Co/Cu(001) (squares). Black lines indicate the pump-probe cross-correlation (CC) measured at the sample surface. It is shown inverted in (a). In addition, solid lines show (a) the relative change of spin-integrated charge carriers $n_{\mathrm{Co}}$ and (b) the relative change of the Co magnetic moment as calculated by TDDFT. Data for different Co thicknesses are vertically offset for display. The inset depicts $\Delta^{2\omega}_{\mathrm{odd}}$ for longer delays.}
    \label{fig:2}
\end{figure}

We theoretically analyze the spin-dependent microscopic processes contributing to laser-induced spin dynamics with parameter-free, fully \textit{ab-initio} time-dependent density functional theory (TDDFT). Our TDDFT calculations were performed for slabs of 3 or 5~ML Co on top of 7~ML Cu(001), and a pump laser pulse of 35~fs (FWHM) pulse length, 800~nm wavelength (1.55~eV photon energy) and $F=0.25$~mJ/cm$^2$. This pump fluence is equal to the fraction of the experimentally employed fluence which is absorbed in the Co/Cu(001) heterostructure. In this way, we properly account for the fact that $\approx 94$~\% of the incident pump fluence is reflected from the sample surface. This could not otherwise be taken into account, as calculating the coupled dynamics of the electronic system and Maxwell's equations is too computationally demanding. As implemented in the ELK code \cite{ELK, krieger2015, TDDFT_parameters} using the full potential linearized augmented-plane-wave method, we treat the time-dependent Kohn-Sham orbitals as two-component Pauli spinors as follows:
\begin{eqnarray}
\label{eq:hamiltonian}
&i&\frac{\partial \psi_j({\bf r},t)}{\partial t}=\left[\frac{1}{2}\left(-i{\boldsymbol \nabla} +\frac{1}{c}{\bf A_{\rm ext}}(t)\right)^2 +v_{s}({\bf r},t) \right. \\
&+& \left.\frac{1}{2c} {\boldsymbol \sigma}\cdot{\bf B}_{s}({\bf r},t) + \frac{1}{4c^2} {\boldsymbol \sigma}\cdot ({\boldsymbol \nabla}v_{s}({\bf r},t) \times -i{\boldsymbol \nabla})\right] \psi_j({\bf r},t) \nonumber
\end{eqnarray}
with ${\boldsymbol \sigma}$ referring to the Pauli matrices. The Kohn-Sham effective potential $v_{s}({\bf r},t) = v_{\rm ext}({\bf r},t)+v_{\rm H}({\bf r},t)+v_{\rm xc}({\bf r},t)$ consists of the external potential $v_{\rm ext}$, the classical electrostatic Hartree potential $v_{\rm H}$, and the exchange-correlation (XC) potential $v_{\rm xc}$, while the Kohn-Sham magnetic field is ${\bf B}_{s}({\bf r},t)={\bf B}_{\rm ext}(t)+{\bf B}_{\rm xc}({\bf r},t)$, ${\bf B}_{\rm ext}(t)$ being the magnetic field of the laser pulse and ${\bf B}_{\rm xc}({\bf r},t)$ the XC magnetic field. We use the adiabatic local spin density approximation for ${\bf B}_{\rm xc}({\bf r},t)$, see \cite{krieger2015}. The last term in Eq.~\ref{eq:hamiltonian} represents spin-orbit coupling. ${\bf A_{\rm ext}}(t)$ is the vector potential representing the pump field. Since we only time-propagate the electronic system while keeping the nuclei fixed (Born-Oppenheimer approximation), our comparison with experiment focuses on the first $\approx 100$~fs after optical excitation. At later times coherent \cite{henighan2016} and incoherent \cite{wieczorek2015, koopmans2009} lattice excitations become important.

\begin{figure}
\includegraphics[width=\columnwidth]{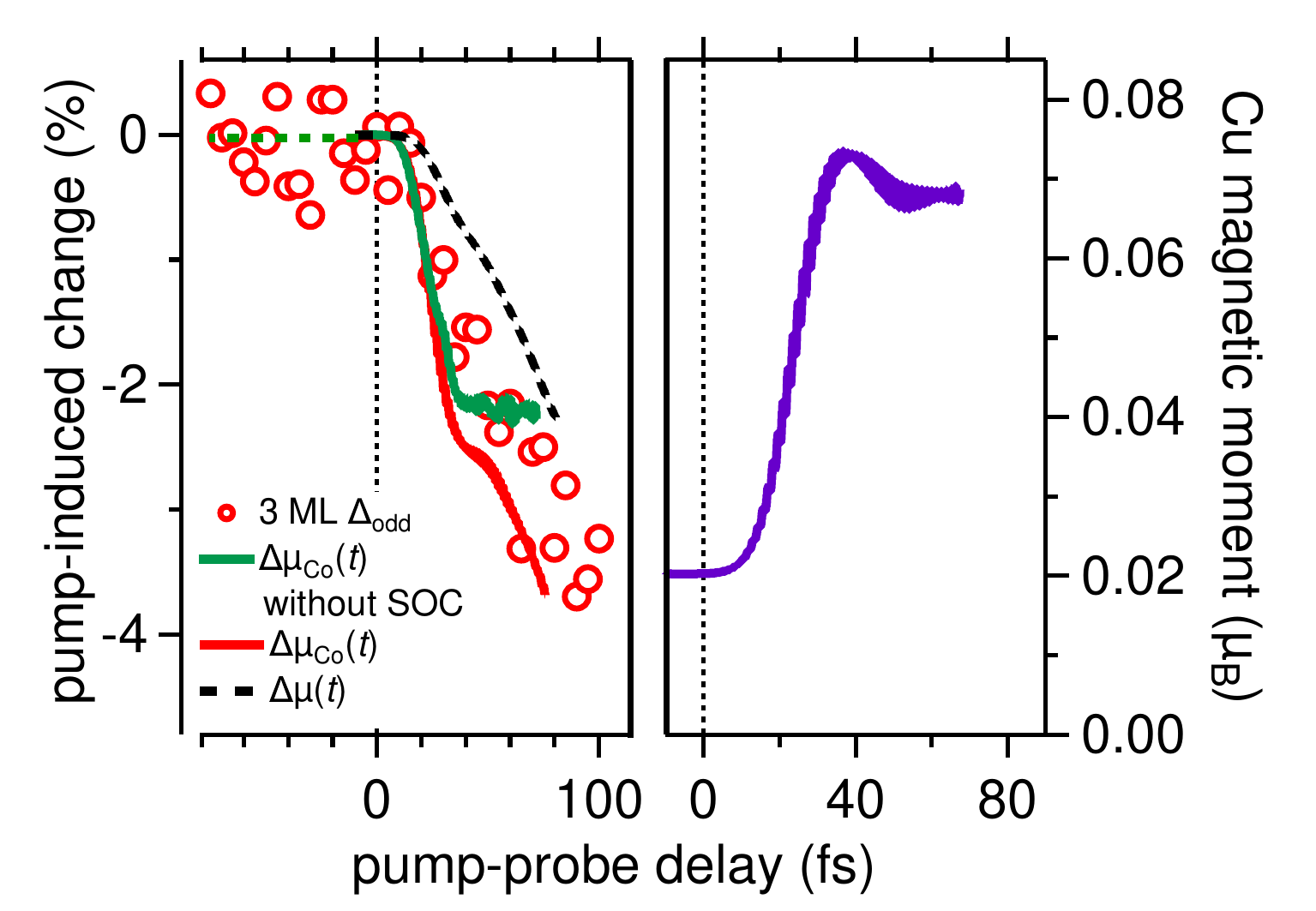}
\caption{\label{fig:3} Left: Comparison of the time-dependent change of magnetic moment for 3~ML Co/Cu(001) as observed in experiment by $\Delta^{2\omega}_{\mathrm{odd}}$ and by TDDFT calculations. The relative changes (normalized to values $\mu(t<0)$ before excitation) of the total spin moment in Co/Cu(001) $\Delta\mu$ as well as of the Co contributions $\Delta\mu_{\mathrm{Co}}$ with and without SOC are given. Right: Magnetic moment induced in Cu $\mu_{\mathrm{Cu}}$ by spin transfer across the interface.}
\end{figure}

Experimental results are shown in Fig.~\ref{fig:2} for 3 and 5 ML~Co/Cu(001), together with the pump-probe cross correlation (CC) measured on the sample surface, which indicates the experimental time resolution. $\Delta^{2\omega}_{\mathrm{even}}$ (Fig.~\ref{fig:2}(a)), which reflects the charge dynamics, exhibits an immediate response to the pump excitation, starting at delay times $t<0$ due to the finite pulse duration and with the maximum change being reached before 50~fs. Within 700~fs, the signal recovers to the value before the pump excitation \cite{supp_mat}. The TDDFT results in Fig.~\ref{fig:2}(a) analyze the relative change in the spin-integrated charge carrier number in the Co film $n_{\mathrm{Co}}(t)$. Since the total charge count is constant, $n_{\mathrm{Co}}(t)$ quantifies the flow of optically excited, spin-integrated charge carriers across the Co/Cu(001) interface, from Co to Cu. $n_{\mathrm{Co}}(t)$ and $\Delta^{2\omega}_{\mathrm{even}}$ show a different evolution with $t$ because TDDFT retrieves the pump-induced charge response directly, while in the experiment a probe pulse is required to monitor the dynamics, which is also sensitive to the coherent polarization induced already by the leading edge of the pump pulse, determining the response to the CC. However, $n_{\mathrm{Co}}(t)$ and $\Delta^{2\omega}_{\mathrm{even}}$ reach their respective minima at the same $t$ within uncertainties. We conclude that the build-up of $\Delta^{2\omega}_{\mathrm{even}}$ is dominated by spin-integrated charge transfer dynamics across the interface. Note that even the subtle difference in the magnitude of $\Delta^{2\omega}_{\mathrm{even}}$ for 3 and 5 ML Co/Cu(001) is reproduced by TDDFT. We further observe a recovery of $n_{\mathrm{Co}}(t)$ after about 30~fs, which indicates a backflow of charge from Cu to Co.

In contrast to $\Delta^{2\omega}_{\mathrm{even}}$, $\Delta^{2\omega}_{\mathrm{odd}}$ (Fig.~\ref{fig:2}(b)) is characterized by a qualitatively different behavior. The change starts at $t=0$ and is slower than $\Delta^{2\omega}_{\mathrm{even}}$, with the minimum of $\Delta^{2\omega}_{\mathrm{odd}}$ being reached at $\approx 100$~fs, much later than the minimum of $\Delta^{2\omega}_{\mathrm{even}}$. This observation hints at additional processes observed in the dynamics of the interface magnetization probed by $\Delta^{2\omega}_{\mathrm{odd}}$. This is further supported by the slower recovery of $\Delta^{2\omega}_{\mathrm{odd}}$ compared to $\Delta^{2\omega}_{\mathrm{even}}$, see the inset of Fig.~\ref{fig:2}(b) and \cite{supp_mat}. The timescale of $\Delta^{2\omega}_{\mathrm{odd}}$ matches the calculated time-dependent change of the Co spin magnetic moment $\Delta\mu_{\mathrm{Co}}(t)=[\mu_{\mathrm{Co}}(t)-\mu_{\mathrm{Co}}(t<0)]/\mu_{\mathrm{Co}}(t<0)$, where $\mu_{\mathrm{Co}} \propto n^{\uparrow}_{\mathrm{Co}} - n^{\downarrow}_{\mathrm{Co}}$ ($\uparrow$ and $\downarrow$ refer to majority and minority carriers, respectively), see Fig. \ref{fig:2}(b). The agreement between experiment and theory allows us to derive additional information that experiment alone does not provide. As discussed below, we identify spin-dependent charge transfer from Co to Cu, including back-transfer from Cu to Co, and separate the loss of magnetic moment in Co mediated by spin transfer and by SOC.

Fig.~\ref{fig:3} (left) compares $\Delta^{2\omega}_{\mathrm{odd}}$ for 3~ML Co/Cu(001) to three different calculated curves, which account for the spin moment change (i) in the full heterostructure $\Delta\mu$ in comparison to the Co contribution $\Delta\mu_{\mathrm{Co}}$ (ii) with and (iii) without SOC being taken into account, i.e. without the SOC term in Eq.~\ref{eq:hamiltonian}. In the latter case, the total magnetic moment $\Delta\mu$ is conserved, but due to spin-dependent charge transfer local moments are redistributed between Co and Cu sites. (ii) and (iii) coincide up to about $\approx35$~fs, which indicates the time range during which spin-dependent charge transfer dominates the magnetization dynamics. A comparison of (ii) and (iii) thus quantifies the extent of SOC mediated demagnetization. Without SOC, the loss in $\mu_{\mathrm{Co}}$ stops after $\approx35$~fs, while including SOC it continues up to 100~fs, which roughly doubles the demagnetization. Our finding implies that spin transfer and SOC contribute to ultrafast demagnetization by a similar extent, but dominate on separate, subsequent timescales. Quantitative agreement between theory and experiment thus allows us to identify a SOC mediated contribution to fs demagnetization, as predicted by theory \cite{zhang2000, toews2015, krieger2015, slab, vishal}.

We moreover observe a larger pump-induced decrease in $\mu_{\mathrm{Co}}(t)$ than in $\mu(t)$ at any given $t$. The difference of $\mu(t)$ and $\mu_{\mathrm{Co}}(t)$ provides the dynamics of the spin moment induced in Cu $\mu_{\mathrm{Cu}}(t)$, shown in Fig.~\ref{fig:3} (right). $\mu_{\mathrm{Cu}}(t)$ reaches its maximum at $t=35$~fs and recedes on longer times. In the same way as the Co demagnetization at $35~\mathrm{fs}<t<100$~fs, this reduction is due to the coupling of both spin channels by SOC. Electron-electron scattering then leads to spin-flip processes and spin moment loss. Based on our TDDFT calculations we estimate the spin injection efficiency from Co to Cu before SOC-mediated spin flips start to dominate by $\frac{|\mu_{\mathrm{Cu}}(35~\mathrm{fs})-\mu_{\mathrm{Cu}}(t<0)|}{ |\mu_{\mathrm{Co}}(35~\mathrm{fs})-\mu_{\mathrm{Co}}(t<0)|}$. We obtain a spin injection efficiency of 40~\% (25~\%) for 3 (5) ML Co/Cu(001).

\begin{figure}
\includegraphics[width=\columnwidth]{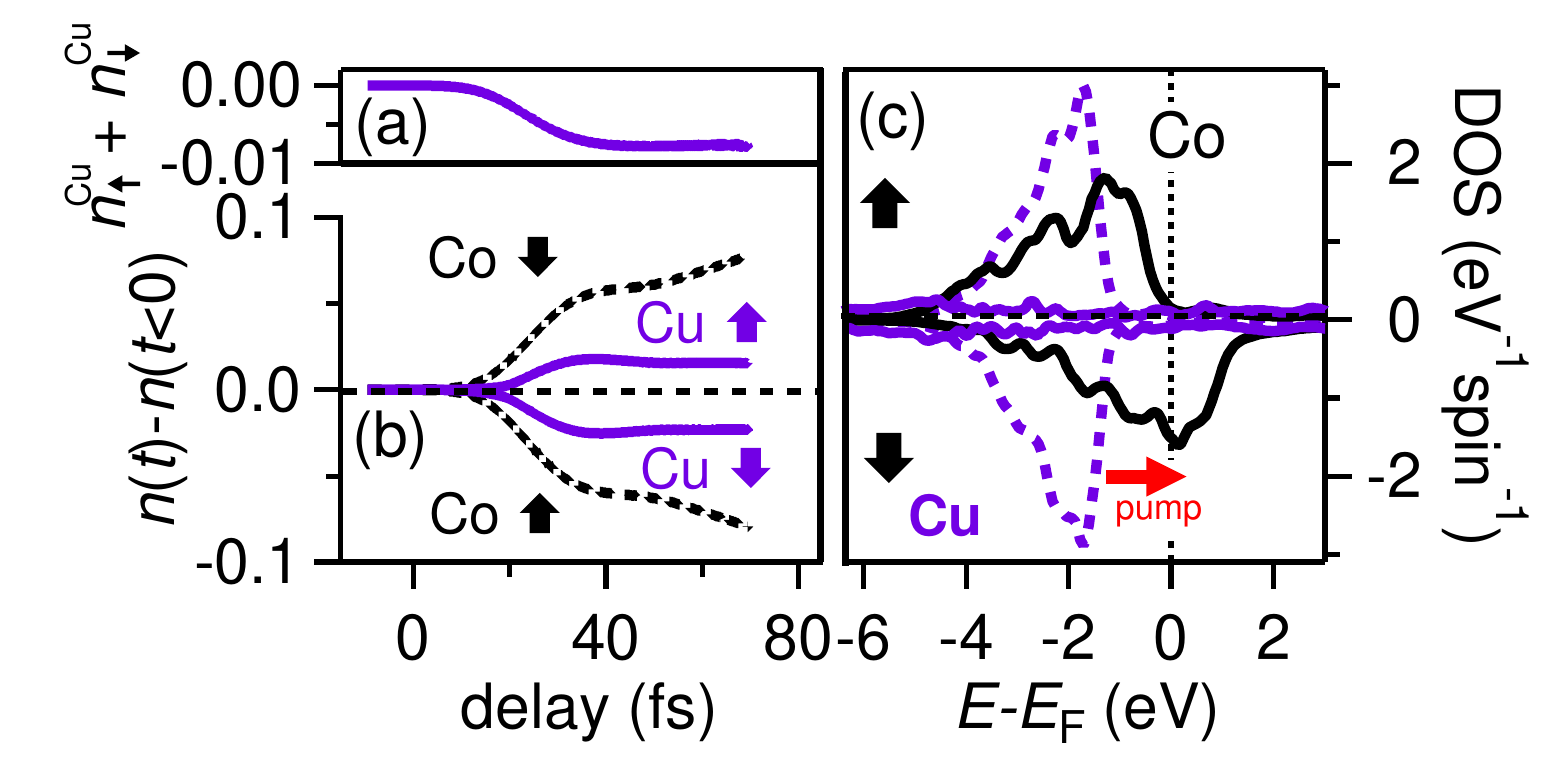}
\caption{(a) Sum of spin polarized carriers in Cu. (b) Calculated change of number of excited majority and minority charge carriers $n^{\uparrow,\downarrow}$ localized at Co and Cu for 3~ML Co/Cu(001) as a function of pump-probe delay. (c) Electronic density of states (DOS) of Co and Cu layers at the interface (Cu $d$ and $sp$ states: dashed respectively solid lines). The horizontal arrow indicates the optical pump transition.}
    \label{fig:4}
\end{figure}

The dynamics of the spin-integrated charge carriers in Cu shown in Fig.~\ref{fig:4}(a) reinforces that the dynamics in the initial $\approx 35$~fs are driven by charge transfer, as the pump-induced change saturates after this time. Fig.~\ref{fig:4}(b) displays the calculated time-dependent change of the number of majority ($\uparrow$) and minority ($\downarrow$) carriers in the Co ($n_{\mathrm{Co}}$) and Cu ($n_{\mathrm{Cu}}$) layers. The increase of $n^{\uparrow}_{\mathrm{Cu}}$ simultaneous with a decrease of $n^{\uparrow}_{\mathrm{Co}}$ is the consequence of spin transfer from ferromagnetic Co to paramagnetic Cu. However, the increase in Cu is much weaker than the decrease in Co, indicating the role of competing SOC mediated spin-flips, which limit the majority spin injection efficiency from Co to Cu. The two different rates of change of $n^{\downarrow}_{\mathrm{Co}}$ and $n^{\uparrow}_{\mathrm{Co}}$ before and after $\approx 35$~fs indicate that the dominant microscopic process changes from spin transfer across the interface to local spin flips mediated by SOC at this point of time. The turning point at $\approx 35$~fs coincides with the pump pulse length, which suggests that spin transfer dominates as long as the pump pulse excites further carriers.

Moreover, we note that $n^{\uparrow}_{\mathrm{Cu}}$ increases more weakly than its counterpart $n^{\downarrow}_{\mathrm{Cu}}$ decreases. This behavior is explained by a back-transfer of minority carriers from Cu to Co, which supports the ultrafast demagnetization in the Co film. We explain this back-transfer by a resonant optical excitation with the employed 1.5~eV pump photon energy from occupied Cu minority $3d$ states to unoccupied Co minority $3d$ states. As depicted in Fig. \ref{fig:4}(c), the electronic density of states (DOS) supports such an optically driven minority spin back-transfer from Cu to Co only directly at the interface, where hybridization of Cu and Co generates new Cu $3d$ states closer to the Fermi energy $E_\mathrm{F}$ than in bulk Cu \cite{nilsson1996, ling2002}. This finding indicates that optically excited spin transfer is determined by the available electronic states around $E_\mathrm{F}$, which can be tuned by choice of substrate and/or pump laser frequency.

Inclusion of spin-dependent charge transfer from the substrate and the actual interface electronic structure distinguishes our \textit{ab-initio} approach from model calculations such as the superdiffusive spin transport model \cite{battiato2010, battiato2012}, which in contrast mainly accounts for spatial transport gradients in films based on spin-dependent lifetimes and velocities in FM only. This Cu to Co back-transfer demonstrates that spin-dependent charge transfer at interfaces can contribute significantly to ultrafast spin dynamics. Therefore, it is not sufficient to consider spin-dependent, hot electron lifetimes as input for modeling laser-induced spin transport, but spin-dependent charge transfer excitation across interfaces contributes in addition. Caution is warranted when spin-dependent lifetimes are analyzed in ultrathin FM films due to possible contributions of non-local spin transfer at interfaces.

We find that SOC-mediated spin flips can occur on sub 100~fs timescales as reported in literature for model calculations without accounting for the excitations of phonons \cite{zhang2000, toews2015}. This timescale is much faster than assumed in models based on Elliott-Yafet-like phonon-mediated spin-flip scattering \cite{koopmans2009}, and might therefore be closer to the rate-limiting process in ultrafast demagnetization. On the other hand, SOC mediates the interaction of the magnetic moment with the lattice, which will inevitably act as a sink for the angular momentum \cite{dornes2018}. 

Our present work indicates that three different microscopic processes dominate ultrafast demagnetization at Co/Cu(001) interfaces during subsequent time intervals, starting with spin transfer and back-transfer at $t<35$~fs, followed by SOC mediated spin-flips during $35~\mathrm{fs}<t<100$~fs, and finally phonon-mediated processes. Phonon excitation can induce further demagnetization \cite{koopmans2009, wietstruk2011} but also relaxation as experimentally observed here, when heat transport from the FM layer into the PM substrate leads to dissipation of excess energy.

In summary, we have disentangled the competing microscopic processes occurring after fs optical excitation of Co/Cu(001) interfaces. From a comparison of \textit{ab-initio} TDDFT and interface-sensitive fs time-resolved SHG, we identify spin transfer between Co and Cu governing the dynamics in the first $35$~fs. Subsequently SOC mediated spin flips reduce the overall spin polarization by dissipation to the lattice within 100~fs, and thus limit the efficiency of spin-polarized charge transfer. Our finding of a minority spin back-transfer from Cu to Co due to a resonant optical transition in the interface layers' DOS opens new possibilities for optical control of spin dynamics on fs timescales via tunable laser pulses.

\begin{acknowledgments}
This work was funded by the German Research Foundation (DFG) through SPP 1840 QUTIF, Grants No. ES 492/1-1 and SH 498/3-1. P.~Elliott thanks the CRC/TRR~227 for funding.
\end{acknowledgments}


\begin{thebibliography}{99}

\bibitem{beaurepaire1996} E. Beaurepaire, J.-C. Merle, A. Daunois, and J.-Y. Bigot, Phys. Rev. Lett. \textbf{76}, 4250 (1996).

\bibitem{kirilyuk2010} A. Kirilyuk, A. V. Kimel, and Th. Rasing, Rev. Mod. Phys. {\bf 82}, 2731 (2010).

\bibitem{kirilyuk2013} A. Kirilyuk, A. V. Kimel, and Th. Rasing, Rep. Prog. Phys. \textbf{76}. 026501 (2013).

\bibitem{stupkiewicz2017} A. A. Stupkiewicz, K. Szerenow, A. Afanasiev, A. Kirilyuk, and A. V. Kimel, Nature {\bf 542}, 71 (2017).

\bibitem{kampfrath2011} T. Kampfrath, A. Sell, G. Klatt, A. Pashkin, S. M\"ahrlein, T. Dekorsy, M. Wolf, M. Fiebig, A. Leitenstorfer, and R. Huber, Nat. Photon. \textbf{5}, 31 (2011).

\bibitem{bossini2016} D. Bossini, S. Dal Conte, Y. Hashimoto, A. Secchi, R.V. Pisarev, Th. Rasing, G. Cerullo, and A.V. Kimel, Nat. Commun. \textbf{7}, 10645 (2016).

\bibitem{malinowski2008} G. Malinowski, F. Dalla Longa, J. H. H. Rietjens, P. V. Paluskar, R. Huijink, H. J. M. Swagten, and B. Koopmans, Nature Phys. {\bf 4}, 855 (2008).

\bibitem{battiato2010} M. Battiato, K. Carva, and P. M. Oppeneer, Phys. Rev. Lett. {\bf 105}, 027203 (2010).

\bibitem{melnikov2011} A. Melnikov, I. Razdolski, T. O. Wehling, E. Th. Papaioannou, V. Roddatis, P. Fumagalli, O. Aktsipetrov, A. I. Lichtenstein, and U. Bovensiepen, Phys. Rev. Lett. {\bf 107}, 076601 (2011).

\bibitem{rudolf2012} D. Rudolf {\it et al.}, Nature Commun. {\bf 3}, 1037 (2012).

\bibitem{turgut2013} E. Turgut {\it et al.}, Phys. Rev. Lett. {\bf 110}, 197201 (2013).

\bibitem{eschenlohr2013} A. Eschenlohr, M. Battiato, P. Maldonado, N. Pontius, T. Kachel, K. Holldack, R. Mitzner, A. F\"ohlisch, P. M. Oppeneer, and C. Stamm, Nat. Mater. \textbf{12}, 332 (2013).

\bibitem{kampfrath2013} T. Kampfrath {\it et al.}, Nature Nanotech. {\bf 8}, 256 (2013).

\bibitem{schellekens2014} A.J. Schellekens, K.C. Kuiper, R.R.J.C. de Wit, and B. Koopmans, Nature Commun. {\bf 5}, 4333 (2014).

\bibitem{razdolski2017} I. Razdolski, A. Alekhin, N. Ilin, J. P. Meyburg, V. Roddatis, D. Diesing, U. Bovensiepen, and A. Melnikov, Nature Commun. {\bf 8}, 15007 (2017).

\bibitem{hofherr2017} M. Hofherr {\it et al.}, Phys. Rev. B {\bf 96}, 100403(R) (2017).

\bibitem{graves2013} C. E. Graves \textit{et al.}, Nature Mater.{\bf 12}, 293 (2013).

\bibitem{rothberg2000} L. Rothberg (ed.), \textit{Photo-induced Charge Transfer}, World Scientific, 2000.

\bibitem{nitzan} A. Nitzan, \textit{Chemical Dynamics in Condensed Phases}, Oxford Univ. Press, 2006.

\bibitem{staehler2008} J. St\"ahler, U. Bovensiepen, M. Meyer, and M. Wolf, Chem. Soc. Rev. \textbf{37}, 2180-2190 (2008).

\bibitem{cinchetti2009} M. Cinchetti, K. Heimer, J.-P. W\"ustenberg, O. Andreyev, M. Bauer, S. Lach, C. Ziegler, Y. Gao, and M. Aeschlimann, Nat. Mater. \textbf{8}, 115 (2009).

\bibitem{chang2015} W. Chang \textit{et al.}, Nature Commun. \textbf{6}, 6415 (2015).

\bibitem{oistr} J. K. Dewhurst, P. Elliott, S. Shallcross, E. K. U. Gross, and S. Sharma, Nano. Lett. \textbf{18}, 184 (2018).

\bibitem{zhang2000} G. P. Zhang and W. H\"ubner, Phys. Rev. Lett. {\bf 85}, 3025-3028 (2000).

\bibitem{toews2015} W. T\"ows and G. M. Pastor, Phys. Rev. Lett. {\bf 115}, 217204 (2015).

\bibitem{wieczorek2015} J. Wieczorek, A. Eschenlohr, B. Weidtmann, M. R\"osner, N. Bergeard, A. Tarasevitch, T. O. Wehling, and U. Bovensiepen, Phys. Rev. B {\bf 92}, 174410 (2015).

\bibitem{jal2017} E. Jal, V. L\'{o}pez-Flores, N. Pontius, T. Fert\'{e}, N. Bergeard, C. Boeglin, B. Vodungbo, J. L\"{u}ning, and N. Jaouen, Phys. Rev. B {\bf 95}, 184422 (2017).

\bibitem{reid2018} A. H. Reid {\it et al.}, Nature Commun. {\bf 3}, 388 (2018).

\bibitem{dornes2018} C. Dornes {\it et al.}, arXiv:1804.07159 (2018).

\bibitem{guedde1999} J. G\"udde, U. Conrad, V. J\"ahnke, J. Hohlfeld, and E. Matthias, Phys. Rev. B \textbf{59}, R6608-R6611 (1999).

\bibitem{regensburger2000} H. Regensburger, R. Vollmer, and J. Kirschner, Phys. Rev. B \textbf{61}, 14716-14722 (2000).

\bibitem{rasing1999} Th. Rasing, Appl. Phys. B \textbf{68}, 477-484 (1999).

\bibitem{schmidt2005} A. B. Schmidt, M. Pickel, M. Wiemh\"ofer, M. Donath, and M. Weinelt, Phys. Rev. Lett. \textbf{95}, 107402 (2005).

\bibitem{melnikov2003} A. Melnikov, I. Radu, U. Bovensiepen, O. Krupin, K. Starke, E. Matthias, and M. Wolf, Phys. Rev. Lett. \textbf{91}, 227403 (2003).

\bibitem{chen2017} J. Chen, J. Wieczorek, A. Eschenlohr, S. Xiao, A. Tarasevitch, and U. Bovensiepen, Appl. Phys. Lett. \textbf{110}, 092407 (2017).

\bibitem{weber1996} W. Weber, A. Bischof, R. Allenspach, C. H. Back, J. Fassbender, U. May, B. Schirmer, R. M. Jungblut, G. G\"untherodt, and B. Hillebrands, Phys. Rev. B \textbf{54}, 4075 (1996).

\bibitem{jaehnke1999} V. J\"ahnke, U. Conrad, J. G\"udde, and E. Matthias, Appl. Phys. B \textbf{68}, 485 (1999).

\bibitem{conrad2001} U. Conrad, J. G\"udde, V. J\"ahnke, and E. Matthias, Phys. Rev. B \textbf{63}, 144417 (2001).

\bibitem{note_even} For several ML thick Co/Cu(001) films, $E_{\mathrm{even}}/E_{\mathrm{odd}}$ is found to be $\approx 10$. Following Ref. \cite{conrad2001}, the SH signal is dominated by a charge response detected in $E_{\mathrm{even}}$ up to 2~ML Co thickness, while the film is at room temperature still paramagnetic. Due to this dominant charge response, a contribution to SHG $\propto E_{\mathrm{odd}}^2$ is neglected after ferromagnetic order sets in at thicknesses $>2$~ML.

\bibitem{ELK} http://elk.sourceforge.net

\bibitem{krieger2015} K. Krieger, J. K. Dewhurst, P. Elliott, S. Sharma, and E. K. U. Gross, J. Chem. Theory Comput. \textbf{11}, 4870 (2015).

\bibitem{TDDFT_parameters} A regular mesh in {\bf k}-space of $8\times8\times4$ and a smearing width of 0.027~eV were used. A time step of 2~as was employed for the time-propagation algorithm. The laser pulse is linearly polarized.

\bibitem{henighan2016} T. Henighan \textit{et al.}, Phys. Rev. B \textbf{93}, 220301 (2016).

\bibitem{koopmans2009} B. Koopmans, G. Malinowski, F. Dalla Longa, D. Steiauf, M. F\"{a}hnle, T. Roth, M. Cinchetti, and M. Aeschlimann, Nat. Mater. {\bf 9}, 259 (2009).

\bibitem{supp_mat} See supplementary material.

\bibitem{slab} K. Krieger, P. Elliott, T. M\"uller, N. Singh, J. K. Dewhurt, E. K. U. Gross, and S Sharma, J. Phys. Condens. Matter \textbf{29}, 224001 (2017).

\bibitem{vishal} V. Shokeen, M. Sanchez Piaia, J.-Y. Bigot, T. M\"uller, P. Elliott, J. K. Dewhurst, S. Sharma, and E. K. U. Gross, Phys. Rev. Lett. \textbf{119}, 107203 (2017).

\bibitem{heusler} P. Elliott, T. M\"uller, J. K. Dewhurst, S. Sharma, and E. K. U. Gross, Sci. Rep. \textbf{6}, 38911 (2016).

\bibitem{nilsson1996} A. Nilsson, J. St\"ohr, T. Wiell, M. Alden, P. Bennich, N. Wassdahl, M. G. Samant, S. S. P. Parkin, N. Martensson, J. Nordgren, B. Johansson, and H. L. Skriver, Phys. Rev. B \textbf{54}, 2917-2921 (1996).

\bibitem{ling2002} W. L. Ling, E. Rotenberg, H. J. Choi, J. H. Wolfe, F. Toyama, S. Paik, N. V. Smith, and Z. Q. Qiu, Phys. Rev. B \textbf{65}, 113406 (2002).

\bibitem{battiato2012} M. Battiato, K. Carva, and P. M. Oppeneer, Phys. Rev. B {\bf 86}, 024404 (2012).



\bibitem{wietstruk2011} M. Wietstruk, A. Melnikov, C. Stamm, T. Kachel, N. Pontius, M. Sultan, C. Gahl, M. Weinelt, H. A. D\"urr, and U. Bovensiepen, Phys. Rev. Lett. \textbf{106}, 127401 (2011).

\end{thebibliography}
\end{document}